\documentclass[aip,jcp,twocolumn,reprint,graphicx]{revtex4-1}

\draft 

\pdfoutput=1
\usepackage{hyperref,url}
\usepackage{amsmath}
\usepackage{amsfonts}
\usepackage[capitalise]{cleveref}
\usepackage{placeins}
\hypersetup{breaklinks,colorlinks=true,linkcolor=blue,citecolor=blue,urlcolor=blue,filecolor=blue}
\usepackage{graphicx}
\usepackage{dcolumn}

\begin{document}


\title{Accelerating the Convergence of Auxiliary-Field Quantum Monte Carlo in Solids with Optimized Gaussian Basis Sets} 



\author{Miguel A. Morales}
\email[]{moralessilva2@llnl.gov}
\affiliation{Quantum Simulations Group, Lawrence Livermore National Laboratory, Livermore, California 94550, USA}

\author{Fionn D. Malone}
\email[]{malone14@llnl.gov}
\affiliation{Quantum Simulations Group, Lawrence Livermore National Laboratory, Livermore, California 94550, USA}


\date{\today}

\begin{abstract}
We investigate the use of optimized correlation consistent gaussian basis sets for the study of insulating solids with auxiliary-field quantum Monte Carlo (AFQMC).
The exponents of the basis set are optimized through the minimization of the second order M{\o}ller--Plesset perturbation theory (MP2) energy in a small unit cell of the solid.
We compare against other alternative basis sets proposed in the literature, namely calculations in the Kohn--Sham basis and in the natural orbitals of an MP2 calculation.
We find that our optimized basis sets accelerate the convergence of the AFQMC correlation energy compared to a Kohn--Sham basis, and offer similar convergence to MP2 natural orbitals at a fraction of the cost needed to generate them.
We also suggest the use of an improved, method independent, MP2-based basis set correction that significantly reduces the required basis set sizes needed to converge the correlation energy.
With these developments, we study the relative performance of these basis sets in LiH, Si and MgO, and determine that our optimized basis sets yield the most consistent results as a function of volume.
Using these optimized basis sets, we systematically converge the AFQMC calculations to the complete basis set and thermodynamic limit and find excellent agreement with experiment for systems studied.
Although we focus on AFQMC, our basis set generation procedure is independent of the subsequent correlated wavefunction method used.
\end{abstract}

\pacs{}

\maketitle 

\section{Introduction}

The accurate theoretical description of complex materials is one of the grand challenges of chemistry and physics.
Mean field methods, such as density functional theory\citep{HohenbergDFT1964,KohnDFT1965} (DFT), are by far the most widely used approaches, offering often satisfactory accuracy relative to their computational cost.
Despite this, DFT suffers from a number of well known challenges, including the choice of approximate exchange correlation functional\citep{Medvedev_functional_2017} and with the somewhat ad-hoc extensions necessary for the treatment of strongly correlated systems\citep{LDAU1998}.
Motivated by these limitations, new methodological developments and by the ever increasing computer power available\citep{KotheECP2020}, there has been a renewed interest in applying wavefunction based quantum chemistry methods in solids\citep{BoothSolids2013}.
Although typically orders of magnitude more expensive than (DFT), wavefunction methods are appealing as they form a systematic hierarchy of approximate approaches that yield a higher accuracy albeit at an ever increasing cost.

Perhaps the most widely developed solid state quantum chemistry methods to date are second-order M{\o}ller--Plesset pertubation theory\citep{AyalaLTMP22001,UsvyatMP22007_a,UsvyatMP22007_b,maschio_2010_pmp2,MarsmanMP22009,GruneisMP22010,delben_2012_pmp2} (MP2) and the Random Phase Approximation (RPA) \citep{Harl09,Harl10,Paier12,Xinguo12}. Coupled cluster methods have also been recently developed and applied to solid state problems \citep{GruneisMgO2015,GruneisCCTDL2018,mcclain2017gaussian} (CCSD). 
In parallel with this, orbital based statistical approaches such as full configuration interaction Monte Carlo\citep{BoothFCIQMC2009,ClelandInitiator2010}
and in particular, auxiliary-field quantum Monte Carlo\citep{zhang_cpmc,Zhang_phaseless}, have begun to be widely applied in solids\citep{BoothSolids2013,PurwantoPressure2009,ma_excited_state_2013,suewattana_pw_phaseless,ma_multiple_proj,zhang_nio,purwanto_downfolding_jctc,ma_downfold_2015,malone_isdf,MaloneGPU2020} (AFQMC).
Despite this rapid development, the widespread adoption of these approaches has in part been hindered by the lack of compact basis sets available. This is critically important if experimentally accurate relative energies are desired as all of these approaches scale with a high power of the basis set size.

Many different approaches have been developed over the years to accelerate the convergence of wavefunction methods in solids with respect to the single particle basis set.
These include constructing `downfolded' Hamiltonians\citep{purwanto_downfolding_jctc,ma_downfold_2015}, using natural orbitals from MP2\citep{GruneisNatOrb2011}, employing explicitly correlated approaches\citep{tew_expl_corr,GruneisF122013,GruneisMgO2015}, developing mixed gaussian-plane-wave approaches\citep{BoothPW2016} and using correction schemes from lower order theories\citep{IrmlerDuality2019,IrmlerMP2Basis2019}.
Despite these developments, no one approach is ideal for every method and converge to the complete basis set limit remains slow.
In addition to the slow convergence with basis set size, approaches based on GTOs optimized for the atomic environment are also known to suffer from linear dependency issues when transferred to the solid\citep{SuhaiLinDep1982,KudinLinearScalingHF2000,HeydHSEGaps2005,UsyvatLMP22011}.
This is particularly a problem when approaching the complete basis set limit in general and when investigating high pressure phases of materials.

Recently, it has been suggested that custom built GTO basis sets tailored for the solid environment often outperform standard basis set libraries for DFT calculations\citep{DagaOptGaussian2020}.
In this paper, we explore a similar approach, but instead targeted towards generating correlation consistent basis sets for correlated methods.
In essence we propose optimizing the exponents through the minimization of the MP2 correlation energy computed in the unit cell of the material.
The resulting basis sets are then used for this system in different conditions, e.g., different supercell sizes, similar structures and at different volumes.
Although our approach is far from ideal as a new basis set needs to be generated for each new application, we argue that this is a price worth paying given that highly accurate approaches are only ever going to be applied in targeted, relatively simple applications for the foreseeable future.
Moreover, the cost of the subsequent correlated calculation almost always dwarfs the cost of generating the basis set itself.

In this paper we focus on deploying these basis sets in AFQMC calculations, but we stress that these basis sets can be used with other highly accurate many-body methods.
By studying a range of insulating solids, we show that these optimized GTOs outperform `downfolded' PBE\citep{PBE1996} orbitals, atomic GTOs, and yield comparable correlation energies to those found using MP2 natural orbitals\citep{GruneisNatOrb2011} at a fraction of the cost.
We also suggest the use of a simple MP2-based basis set correction scheme to further accelerate convergence that is independent of the underlying wavefunction method in contrast to that suggested in Ref.~\citenum{IrmlerMP2Basis2019}.
With these developments we are able to reliably produce AFQMC results for the energy in the complete basis set and thermodynamic limit for LiH, Si and MgO.
Our results for the bulk modulus, lattice constant and cohesive all agree excellently with experiment.

This rest of this article is outlined as follows. In \cref{sec:basis_sets} we outline our method for generating optimized basis sets for correlated calculations in solids. In \cref{sec:mp2_corr} we describe our MP2-based basis set correction and analyze its performance in a range of settings. In \cref{sec:results} we compare the performance of the various basis sets and present results for the equilibrium properties of LiH, Si and MgO.
Finally, we close in \cref{sec:conclusions} with some discussion on the remaining challenges, including the adaptation of our approach to metals, and the development of better finite size corrections for orbital based approaches.

\section{Basis set generation schemes\label{sec:basis_sets}}

Orbital-based quantum many-body methods are typically formulated starting from the Hamiltonian in second quantization,
\begin{align}
    \hat{H} &= \sum_{ij\sigma} h_{ij} \hat{c}^{\dagger}_{i\sigma}\hat{c}_{j\sigma} + \frac{1}{2}\sum_{ijkl,\sigma{\sigma}^\prime}v_{ijkl}  
 \hat{c}^{\dagger}_{i\sigma}\hat{c}^{\dagger}_{j\sigma^\prime}\hat{c}_{l\sigma^\prime}\hat{c}_{k\sigma},\label{eq:hamil}\\
            &= \hat{H}_1 + \hat{H}_2,
\end{align}
where $\hat{c}^{\dagger}_{i\sigma}$  and $\hat{c}_{i\sigma'}$ represent fermionic creation and annihilation operators in a given single particle orbital (SPO) set, $h_{ij}$ and $v_{ijkl}$
are the one- and two-electron matrix elements of the Hamiltonian.

The development of appropriate basis sets for electronic structure calculations has a long and rich history in both quantum chemistry and condensed matter communities.
This is particularly true in the case of molecular calculations where the development of optimized basis sets has been extended to almost all aspects of the electronic structure calculation including: electron correlation\cite{DunningCC11989,DunningEA1992,DunningCC31993}, density fitting\cite{BaerendsDF11973,WhittenDF21973,WhittenDF31974}, F12 and other explicitly correlated approaches\cite{KlopperF121987,KutzelniggF121985,HattigF122012,HattigF12CCSD2010,WernerF122007}, core-valence correlation \cite{WoonCoreValence1995}, molecular properties\citep{WoonAugBasis1994}, relativistic effects\cite{BalabanovRelativityCC2005} among many others. On the other hand, for periodic calculations in condensed matter systems, the work has mainly focused so far on the development of good basis sets for mean field methods like DFT.
Due to the low cost of DFT calculations in comparison with typical many-body methods like MP2, RPA, CCSD or AFQMC, the development of DFT basis sets are typically focused on efficiency of computation, robustness in convergence and generic applicability.
Typical choices for basis sets in this case include plane waves\citep{IhmTotalEnergy1979,PayneIterative1992}, periodic versions of the linear combination of atomic orbitals approach\citep{PisaniPBCHF1980,CrystalCode2005,joost_gth_2007,gth_cp2k_2013}, wavelets\citep{GenoveseWavelets2008}, among others.

While the above approaches can work out of the box for any correlated wavefunction approach, they all suffer from different drawbacks.
Ideally we would employ a plane-wave approach, which is perhaps the most natural basis for simulating periodic systems and would allow us to build on decades of experience and developments. In addition, using plane waves would give us access to robust and efficient codes\citep{QE2009,KressePW1996,VASP,CASTEP2005} as well as highly developed databases of pseudopotentials\citep{VanderbiltUSP1990} and Projector Augmented Wave potentials\citep{BlochlPAW1994,KressePAW1999}.
Unfortunately, despite offering a robust and straightforward way to converge the total energy with respect to basis set size, a pure plane-wave approach would be computationally demanding given the large planewave cutoffs required to converge the electron-electron cusps\citep{ShepherdMBWFN2012}.
This is crucially important as AFQMC calculations scales like $\mathcal{O}{[M^3]}-\mathcal{O}{[M^4]}$ (with a large prefactor), where $M$ is the number of single particle basis functions.
One method to overcome this is `downfolding'\citep{purwanto_downfolding_jctc,ma_downfold_2015} Kohn--Sham or Hartree--Fock states from a prior plane-wave calculations, i.e., working in an active space of eigenstates of the prior mean field calculation.
As we will see, this approach typically converges slowly with the number of Kohn--Sham states and relies on good error cancellation, which may be difficult to achieve uniformly as conditions are changed, for example by changing volumes in an equation of state calculation.

The use of localized basis sets for periodic calculations also has a long history\citep{PisaniPBCHF1980}, which has mostly been focussed on mean field approaches like Hartree--Fock and DFT\citep{KudinLinearScalingHF2000,CrystalCode2005,gth_cp2k_2013}.
In this case, typically one employs custom built basis sets for solid state DFT simulations\citep{joost_gth_2007} or periodized atomic GTO basis sets\citep{KudinLinearScalingHF2000,HeydHSEGaps2005}.
Unfortunately, basis sets optimized for DFT simulations are not designed for correlated calculations.
In particular they do not follow the usual angular momentum hierarchy developed for atoms and molecules and thus cannot be used to reliably extrapolate to the CBS limit\citep{MaloneGPU2020}.
Although it is possible to modify these basis sets with basis functions from atomic Dunning basis sets\citep{delben_2012_pmp2,MaloneGPU2020}, this procedure is error prone and suboptimal.
On the other hand, directly using all-electron correlation consistent basis sets developed for atoms and molecules is also problematic given the existence of diffuse functions.
These basis functions typically lead to linear dependency issues making the initial mean field simulation difficult to converge.
The common strategy of removing basis functions with small exponents often helps, however it can lead to subsequent difficulties with basis set extrapolation.
These issues are well known and a number of alternative approaches have been developed, including using local-correlation methods\citep{UsyvatLMP22011} or natural orbitals from MP2\citep{GruneisNatOrb2011} (MP2-NO) or RPA\citep{RambergerRPANO2019} (RPA-NO).

\begin{figure}
\includegraphics[width=\columnwidth]{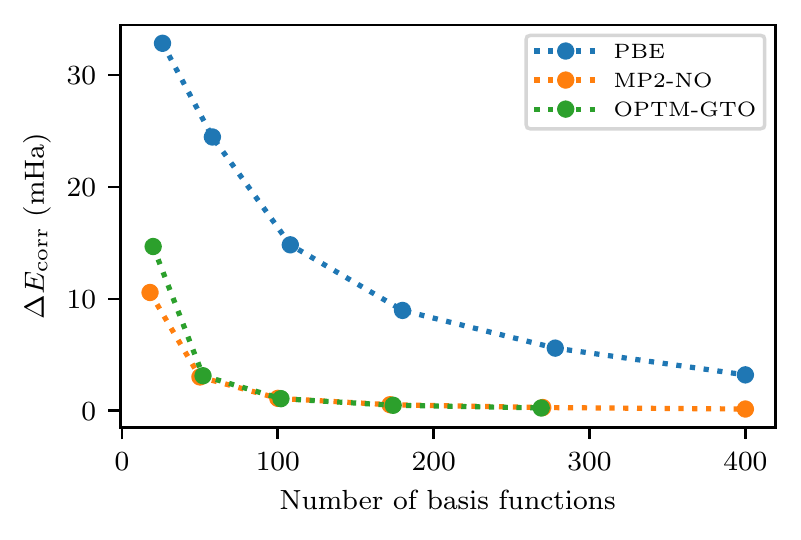}
\caption{Comparison in the convergence in the MP2 correlation energy with PBE, MP2-NO and OPTM-GTO orbitals for a primitive cell of LiH in the B1 structure at equilibrium. The number of PBE and MP2-NO states were chosen to match the number of basis functions in the cc-pVXZ hierarchy (here X=2-6). \label{fig:lih_mp2}}
\end{figure}

Here we seek a practical middle ground, that allows us to use existing plane-wave methodologies while wedding these with more compact localized basis functions designed from the outset for correlated solid state simulations.
Our guiding principles are as follows, we desire a scheme that: 1) has general applicability, 2) can be applied independently of the underlying basis used in the mean field calculations, 3) maximizes the amount of correlation energy for a give basis set size, and 4) lends itself to corrections from lower levels of theory like MP2 or the RPA.
We build on the hybrid PW-PGTO approach recently developed by Booth, \emph{et al.} \cite{BoothPW2016} by not relying on a pre-existing gaussian basis set for the virtual orbitals, instead we optimize the exponents in the basis (and possibly the contraction coefficients if desired) by minimizing the MP2 energy of a small representation of the periodic solid, typically a single primitive unit cell.  Similar to the approach developed by Booth, \emph{et al.}, the current approach allows us to use a fully converged set of occupied orbitals in the reference calculation which improves the transferability of the basis across multiple environments and also allows us to use any representation of the electron-ion interaction without the need to have the PGTO basis involved in the reference calculation. As we will see, the minimization of the MP2 energy leads to a set of virtual states that efficiently captures electron correlation, while at the same time avoiding most of the problems associated with the use of atomic basis sets in solids, like the ubiquity of small exponents which result in catastrophic linear dependency problems.

Our approach to generate a basis set is as follows:
\begin{enumerate}
    \item We first perform a converged PW DFT or HF calculation, to compute the occupied states.
    \item We take the valence states from an existing correlation consistent atomic basis set for the atomic species under consideration and generate the appropriate PGTO orbitals on a FFT grid.
    \item We orthogonalize this PGTO basis against the occupied orbitals.  This set of occupied and virtual orbitals now makes up our basis set.
    \item We next perform a HF calculation within this new set of orbitals.
    \item We compute the MP2 energy from the resulting HF eigenvalues and eigenvectors.
    \item Iterate steps 2-5 by updating all the exponents for the valence states until the MP2 energy is minimized.
\end{enumerate}

As we are looking for a relatively efficient and computationally inexpensive approach, we only reoptimize a basis set for an atom when the environment changes significantly.
This means that in calculations within similar phases, we reuse the basis set generated at one volume for all volumes and system sizes.
In the work presented here, all basis sets were optimized for a single primitive unit cell at the equilibrium volume of the material and reused everywhere.
We also only considered single gaussians without contraction for the virtuals.
In a future publication we will investigate the transferability of the resulting basis for different phases, different combination of elements and as well as other optimization strategies including the use of contracted gaussians.
For now, we are interested in comparing this approach with two other popular and general approaches developed so far, namely downfolded Kohn--Sham states and MP2-NOs.
We don't compare against RPA-NOs, but we expect similar performance in the materials studied here.

As an example of the utility of our approach, in \cref{fig:lih_mp2} we show the error in the MP2 correlation energy, measured with respect to an MP2 calculation using all the plane-wave orbitals within a 70 Ry cutoff, as a function of the number of single particle orbitals of a 2 atom primitive cell of LiH in the B1 structure at the equilibrium volume.
Results are shown for PBE orbitals, MP2-NOs and the MP2-optimized hybrid PW-PGTO basis (from now on referred to as the optimized GTO basis (OPTM-GTO) for simplicity).
As observed by others in the case of LiH \cite{GruneisNatOrb2011,BoothPW2016}, the convergence of the MP2 energy with mean-field orbitals, either KS states like PBE or HF orbitals (not shown here), is extremely slow compared to the convergence on the MP2 natural orbital basis.
While convergence on the latter basis set is always expected to be faster, the different rate of convergence on LiH is quite dramatic.
The convergence of the optimized GTO basis is comparable to the MP2 natural orbital basis, only showing a larger error in the correlation energy for the cc-pVDZ basis.
In this case, it appears that either option represents a reduction of at least a factor of four in the number of states needed to reach a given accuracy, which would imply reductions of over an order of magnitude in AFQMC calculations at the same accuracy.

\section{MP2-based Corrections\label{sec:mp2_corr}}

\begin{figure}
\includegraphics[width=\columnwidth]{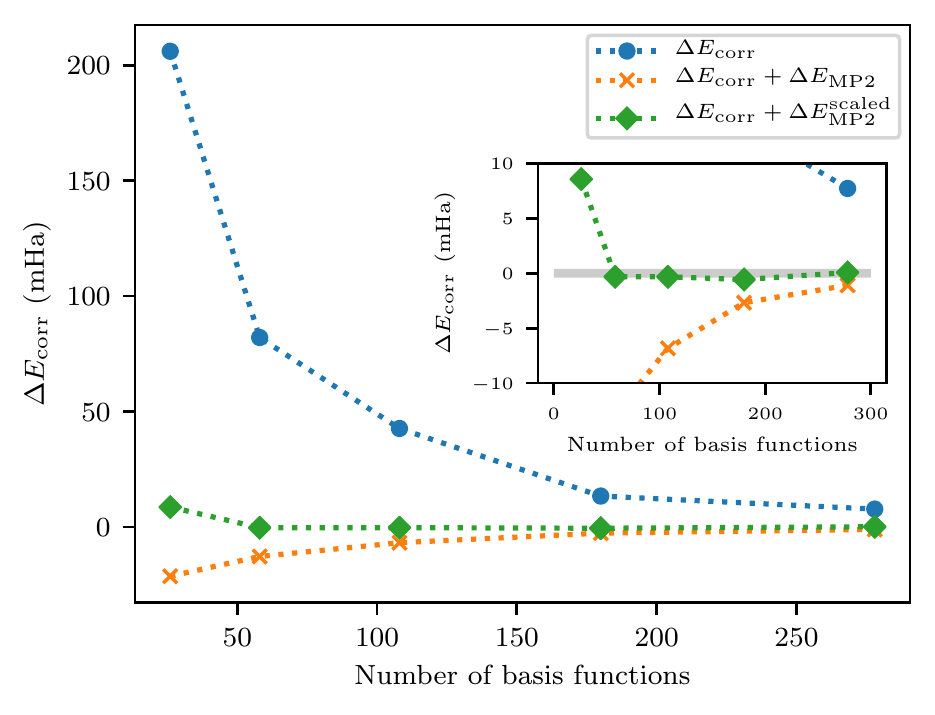}
\caption{Comparison between the convergence of the AFQMC correlation energy with basis sets size when using different MP2-based basis set correction schemes. Here we plot $\Delta E_\mathrm{corr}(M)=E_\mathrm{corr}^{\mathrm{CBS}}-E_\mathrm{corr}(M)$, where $M$ is the number of basis functions. The CBS value has been estimated from a $5,6$ extrapolation using the OPTM-GTO basis sets. The uncorrected AFQMC correlation energy (blue circles) converges slowly when compared to either simple MP2 corrections ($\Delta E_{\mathrm{MP2}}$, orange crosses) or a scaled MP2 correction ($\Delta E_{\mathrm{MP2}}^{\mathrm{scaled}}$, green diamonds). See main text for definitions of the corrections. The inset shows a close up of the convergence demonstrating that the scaled MP2 corrections greatly outperform the simple MP2-based corrections. The system shown here is MgO under ambient conditions in the B1 phase. \label{fig:mp2corr}}
\end{figure}

In addition to fast convergence, another important component of an efficient basis set for correlation is its capacity to provide a framework for basis set corrections, either through extrapolations or in combination with less expensive methods. While basis set extrapolations are very successful in molecular calculations, its utility in condensed matter systems is still unclear. In fact, without properly developed correlation-consistent basis sets for solids this is in general not possible\citep{MaloneGPU2020}. Instead, we will rely on less expensive approaches that offer a basis set dependence that is compatible with AFQMC. In this work we explore the use of MP2 for such a purpose. As is the case with basis set development, the concept of basis set corrections based on MP2 for more expensive many-body methods like CCSD(T) has a long history in the quantum chemistry community\cite{AllenMP2Corr1993,CsaszarMP2Corr1998,JureckaMP2Corr2002,SherrillMP2Corr2009} and its extension to condensed systems is straightforward.  

The simplest MP2 correction adds the difference between the MP2 energy at the complete basis set limit and its value at the same basis used in the AFQMC calculation, namely $\Delta{E_{\mathrm{MP2}}(M)} = E^{\mathrm{CBS}}_{\mathrm{MP2}} - E_{\mathrm{MP2}}(M)$, where $M$ is the number of single particle basis functions.
This correction should be reasonable as long as MP2 is applicable to the system, since MP2 in known to diverge for metallic systems and will be inaccurate for systems with significant static correlation.
Overall, we can expect this correction to overestimate the basis set incompleteness error of AFQMC, since MP2 tends to overestimate the correlation energy in general.

An improved correction can be obtained following the recent work of Irmler and Gruneis \cite{IrmlerDuality2019,IrmlerMP2Basis2019} on MP2 basis set corrections for coupled cluster theory. 
In their work they show that the basis set incompleteness error in CCSD theory is dominated by the MP2 correlation energy and the particle-particle ladder term in the diagrammatic expansion of the CCSD equations, which led them to propose the following correction for CCSD \cite{IrmlerDuality2019}:
\begin{equation}
\Delta{E(M)} = E^{\mathrm{CBS}}_{\mathrm{MP2}} - E_{\mathrm{MP2}}(M) + \left ( \frac{E^{\mathrm{CBS}}_{\mathrm{MP2}}}{E_{\mathrm{MP2}}(M)}-1 \right ) E^{\mathrm{ppl}}(M), \label{eq:ppl}
\end{equation}
where $E^{\mathrm{ppl}}$ is the energetic term associated with the particle-particle ladder (ppl) diagram \footnote{Irmler and Gruneis typically refer to the MP2 energy on their equations as the Driver term.}. Even though the equation is approximate and based on somewhat empirical observations, Irmler and Gruneis subsequently showed that it can be used successfully to significantly reduce basis set incompleteness errors in CCSD \cite{IrmlerMP2Basis2019}. Looking at  \cref{eq:ppl} more closely we notice that 2 approximations are made: 1) the MP2 and ppl terms have the same dependence with basis set size and 2) the rest of the CCSD energy is independent of basis set.

Following similar approximations in this case for the AFQMC energy, we can write:
\begin{equation}
E_{\mathrm{AFQMC}}(M) = \alpha E_{\mathrm{MP2}}(M) + E_0, \label{eq:linearapp}
\end{equation}
where $\alpha$ is a system dependent constant that accounts for the part of the energy that scales like $E_{\mathrm{MP2}}(M)$ and $E_0$ represents those terms which we assume to be insensitive to basis set due to their weak dependence. With this in mind, we can rewrite  \cref{eq:ppl} as:
\begin{equation}
\Delta{E_{\mathrm{MP2}}^{\mathrm{scaled}}(M)} = \alpha \left ( E^{\mathrm{CBS}}_{\mathrm{MP2}} - E_{\mathrm{MP2}}(M) \right ), \label{eq:scaledmp2corr}
\end{equation}
which depends only on $\alpha$ and can be seen as a scaled version of the original MP2 correction. If we compare this expression with  \cref{eq:ppl} in the case of CCSD, we can identify $\alpha = 1 + \frac{E^{\mathrm{ppl}}}{E_{\mathrm{MP2}}}$ which can be directly evaluated at any given point. In AFQMC, unfortunately, we do not have a clear way to calculate it directly.
Instead, we use \cref{eq:linearapp} and solve for $\alpha$ and $E_0$ given the value of the AFQMC and MP2 energies for 2 different vales of $M$.
In the rest of the paper, we refer to  \cref{eq:scaledmp2corr} as the $``$scaled MP2 correction", where $\alpha$ is calculated using results from pVTZ and pVQZ basis sets (or the equivalent number of states with PBE and MP2-NO orbitals) unless otherwise stated.

As an example of this, \cref{fig:mp2corr} shows the error in the AFQMC correlation energy of MgO as a function of the number of orbitals in the OPTM-GTO basis, for a single primitive cell of the solid in the B1 phase. The CBS limit of the AFQMC energy was obtained from an extrapolation using the results from the cc-pV5Z and cc-pV6Z basis sets. The figure shows results for the uncorrected AFQMC energies as well as those with the two corrections discussed above, the unscaled and scaled MP2 corrections. A single value of $\alpha$ was used, calculated using energies from pVTZ and pVQZ basis sets. It is clear that at least for this material, the scaled correction is superior to the bare correction, providing results within error bars of the extrapolated CBS energy already with a cc-pVTZ basis. While this is a particularly successful application of the correction, below we show evidence that the scaled version is in general a superior correction for basis set incompleteness error.

\section{Results\label{sec:results}}

In this section we compare the performance of PBE, MP2-NO and OPTM-GTO orbitals in the calculation of the equation of state of LiH, Si and MgO near equilibrium.
We used a locally modified version of Quantum Espresso\citep{QE2009} to perform HF, DFT and MP2 calculations (including the generation of MP2 natural orbitals), as well as to generate the one- and two-electron Hamiltonian matrix elements for AFQMC. PySCF \cite{SunPYSCF2017,SunPYSCF2020} was used to evaluate the PGTO orbitals in the FFT grid. We used optimized norm-conserving pseudo potentials\cite{HamannONCV2013,SchlipfONCV2015} for all atoms. As described in \cref{sec:basis_sets}, we generated MP2-optimized PGTO basis sets for all atoms by minimizing the MP2 energy of a single primitive unit cell of each material at the equilibrium volume, this basis set was then reused for all volumes and systems sizes presented here. The optimized basis sets have been included in the supplementary material \cite{supplement}. We used the phaseless AFQMC\citep{Zhang_phaseless} method as implemented in QMCPACK\cite{KimQMCPACK2018,KentQMCPACK2020}. We used the hybrid method for weight updates\citep{PurwantoPressure2009} and used the pair-branch population control algorithm\citep{wagner_qwalk}. All AFQMC simulations used a single determinant restricted Hartree--Fock trial wavefunction. The AFQMC input and output data is available at Ref.~\citenum{materials_data}.

\subsection{Basis set convergence}

\begin{figure*}
    \includegraphics[width=\textwidth]{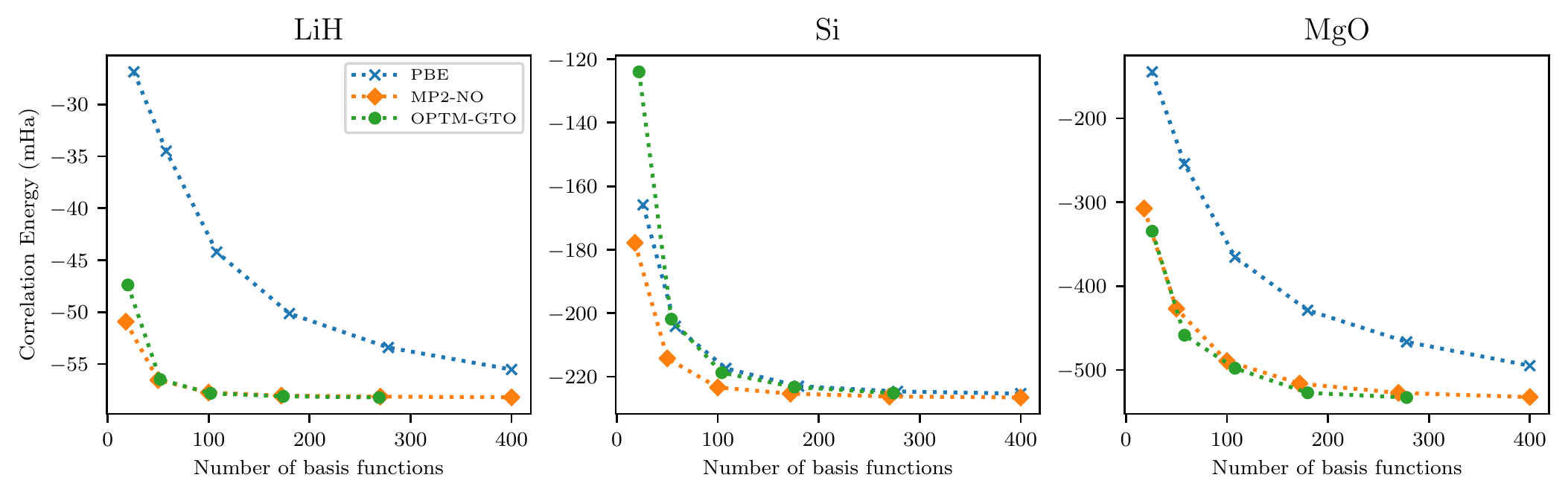}
\caption{Convergence of the AFQMC correlation energy using `downfolded' PBE orbitals, MP2 natural orbitals (MP2-NO) and optimized gaussian orbitals (OPTM-GTO) for unit cells of LiH, Si and MgO in their equilibrium structures. \label{fig:e_vs_N}}
\end{figure*}

 \cref{fig:e_vs_N} shows the AFQMC correlation energy for LiH, Si and MgO, for a primitive unit cell of the material at the equilibrium volume, as a function of the number of orbitals in the single particle basis set. Results are shown for PBE, MP2-NO and OPTM-GTO orbitals in all cases. The dependence of the correlation energy on the basis set size will show weak dependence to system size (e.g. as a function of number of orbitals per atom), which makes calculations on a single primitive cell a useful tool in determining the quality and convergence rate of a given basis. In the case of LiH and MgO, both MP2-NO and OPTM-GTO offer a far superior alternative to downfolded PBE orbitals. We observe a similar performance in these two cases from HF orbitals and from DFT orbitals with other typical exchange-correlation functionals like LDA and PBE0, implying that orbitals obtained from mean-field SCF calculations offer a poor choice on these materials. In the case of Silicon, all three orbitals offer an acceptable convergence rate, with slightly superior results from MP2-NO. We believe that the somewhat slower convergence rate of the OPTM-GTO orbitals is the result of using uncontracted periodic gaussians in the basis. We are currently exploring the use of an optimized contracted basis to further improve performance, which we will report in a future publication.

\begin{figure*}
    \includegraphics[width=\textwidth]{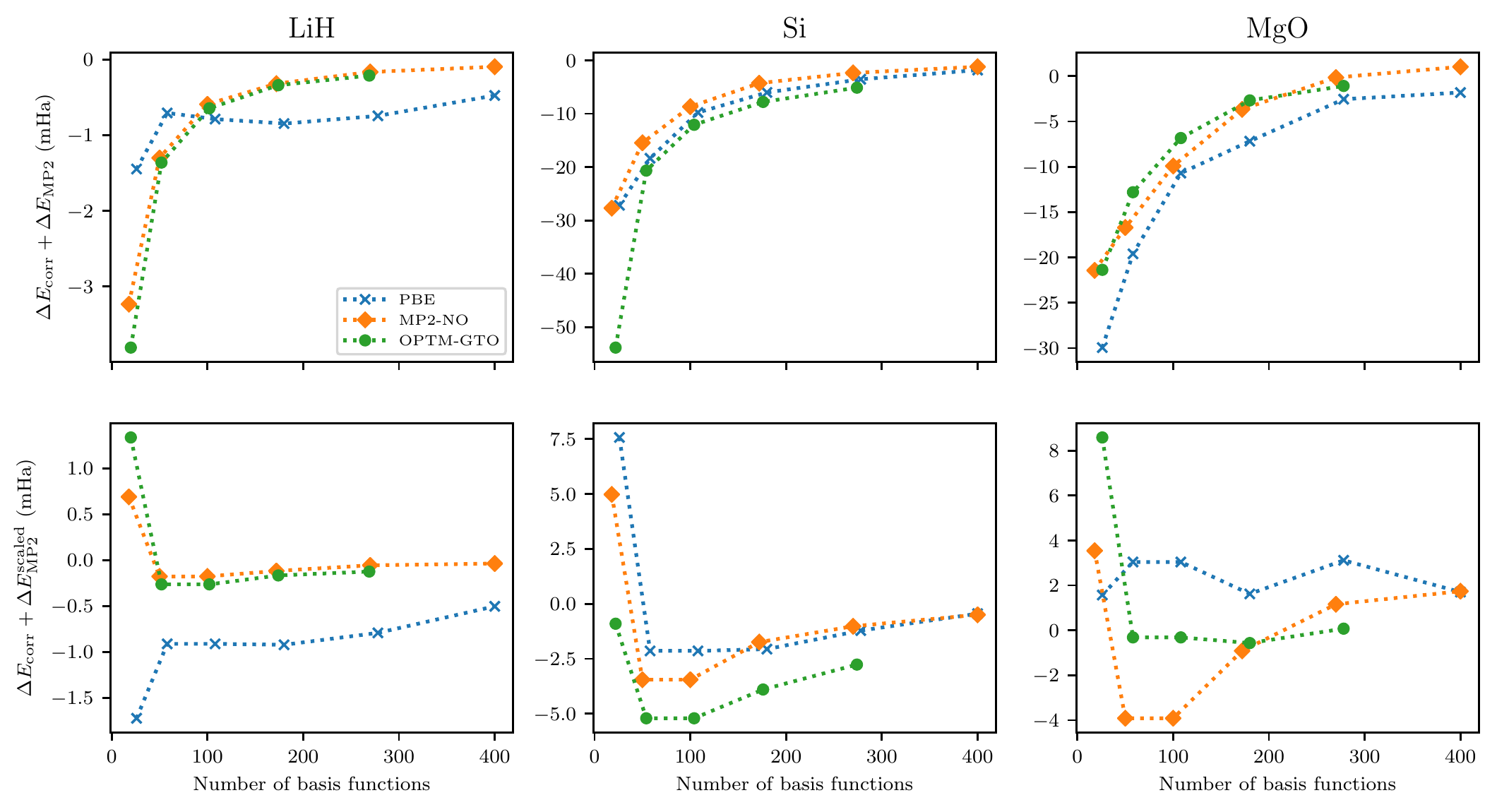}
\caption{Convergence of the AFQMC correlation energy after the MP2 basis set corrections have been applied using `downfolded' PBE orbitals, MP2 natural orbitals (MP2-NO) and optimized gaussian orbitals (OPTM-GTO) for unit cells of LiH, Si and MgO in their equilibrium structures. \label{fig:ecorr_vs_N}}
\end{figure*}

Even though MP2-NO and OPTM-GTO orbitals seem to provide a fast convergence of the correlation energy, the high cost of correlated calculations in solids combined with the need to account for finite size effects, a highly expensive computational task, makes the development of basis set correction schemes a high priority.  \cref{fig:ecorr_vs_N} shows the error in the AFQMC correlation energy, for the same systems shown in  \cref{fig:e_vs_N}, but with the inclusion of the $\Delta{E_{\mathrm{MP2}}}$ and $\Delta{E^{\mathrm{scaled}}_{\mathrm{MP2}}}$ corrections described in \cref{sec:mp2_corr}. In all cases, the scaling factor in $\Delta{E^{\mathrm{scaled}}_{\mathrm{MP2}}}$ was obtained with the results of pVTZ and pVQZ basis sets (or the equivalent number of orbitals in PBE and MP2-NO). Results are plotted with respect to extrapolations to the complete basis set limit from either MP2-NO results (LiH and Si assuming a $1/M$ dependence) or from OPTM-GTO results (for MgO assuming a typical $1/L^3$ dependence). As expected from the discussion in \cref{sec:mp2_corr}, MP2's tendency to overestimate the correlation energy will generally lead to a correction that is too large. The scaled MP2 correction shows a better performance, leading to errors on the order of a few mHa already for pVTZ basis sets. All corrections vanish in the complete basis set limit, leading to the same result at large enough basis set size (this is not always clear with PBE orbitals due to the slow convergence). 
Nonetheless, both corrections offer a significant reduction in the basis set incompleteness error, particularly for the smaller basis sets where the scale of errors can be reduced by over an order of magnitude. In addition, both corrections are able to reduce the error in calculations with PBE orbitals to magnitudes comparable to the other 2 orbital sets. We, nonetheless, still prefer orbitals that directly capture a larger amount of correlation, since with PBE orbitals the correction will be larger than the AFMQC correlation energy with smaller orbital sets.

\subsection{Cold curves and error cancellation}

\begin{figure*}
    \includegraphics[width=\textwidth]{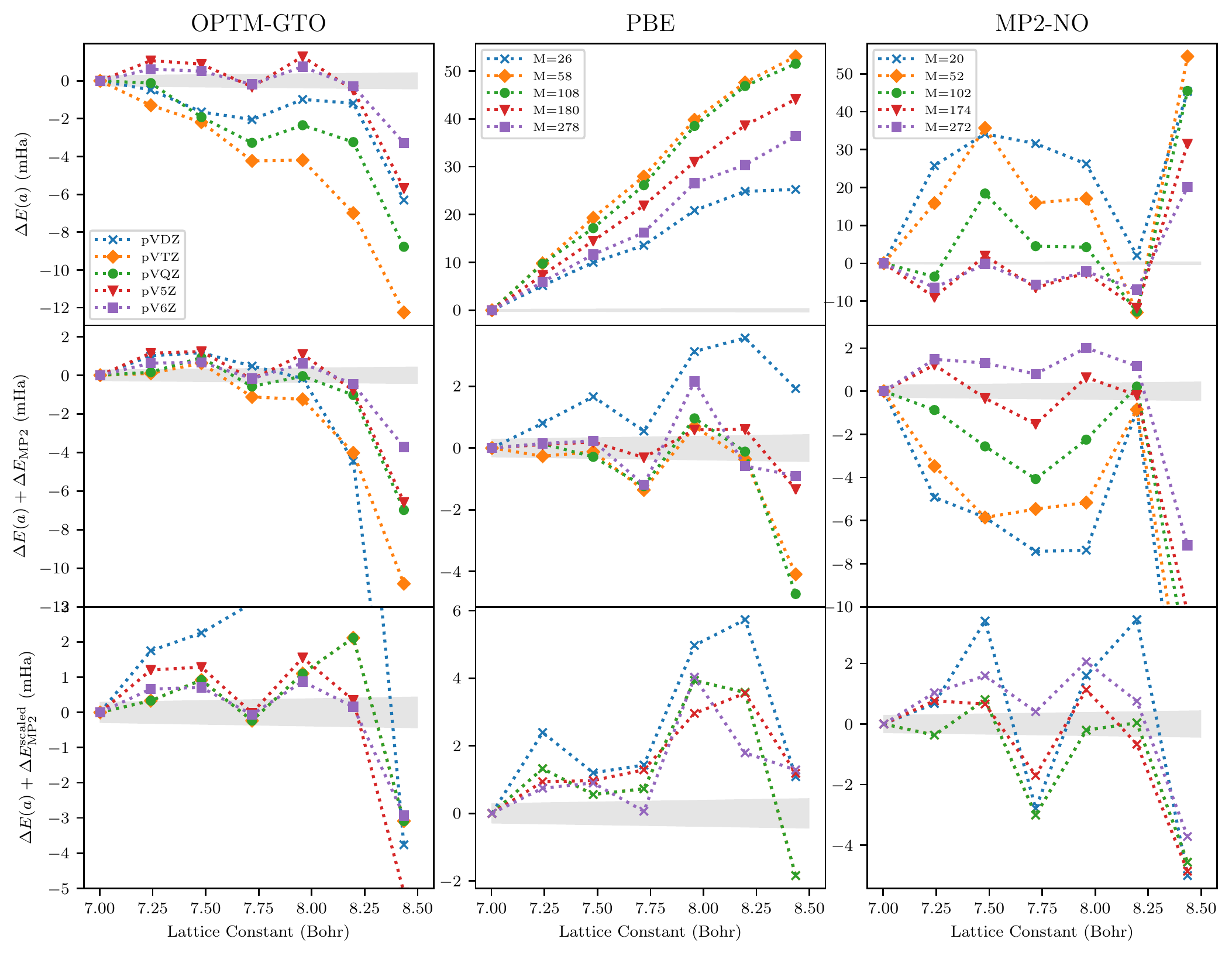}
\caption{Comparison between the calculated non-parallelity errors in the correlation energy ($\Delta E(a))$ found in MgO for the three orbitals sets considered in this work (OPTM-GTO, PBE and MP2-NO). The first row shows the error in the uncorrected AFQMC correlation energy, while the middle and bottom rows show the reduction in error found when using either the standard MP2 correction ($\Delta E_{\mathrm{MP2}}$) or the scaled MP2 correction ($\Delta E_{\mathrm{MP2}}^{\mathrm{scaled}}$) respectively.  The shaded grey region represents the error in the CBS extrapolation.\label{fig:e_vs_a_mgo}}
\end{figure*}

The comparison of these basis sets at a single volume shows similar performance between MP2-NO and OPTM-GTO orbitals, which are both typically superior to mean-field SCF orbitals. If we are interested in calculations at a single volume, both options seem perfectly acceptable. When we are interested in thermodynamic properties of the material that depend on volume, like lattice constants, bulk moduli or the equation of state in general, the more relevant question is not the rate of convergence of the energy at a given volume but rather the cancellation of errors between different volumes. We are interested in a basis set that can approach the correct volume dependence of the correlation energy with as few terms as possible. In the quantum chemistry community these are known as nonparallelity errors in the case of dissociation curves of molecular complexes. 

\cref{fig:e_vs_a_mgo} shows the AFQMC correlation energy of MgO as a function of the lattice constant (the experimental lattice constant, corrected for zero-point effects, is 7.916 Bohr) for all basis sets and varying number of orbitals. In order to better visualize the nonparallelity errors we plot the error in the correlation energy, measured with respect to the complete basis set extrapolated value at every volume, with respect to its value at the lowest volume, namely $\Delta{E(a)} = (E(a)-E^{\mathrm{CBS}}(a))-(E(a_\mathrm{min})-E^{\mathrm{CBS}}(a_\mathrm{min}))$. As currently presented, the plotted value is a combination of the errors at the current and reference volumes. Nonetheless, it clearly shows the magnitude of nonparallelity errors and allows for a clear comparison between various orbital sets. In addition to bare AFQMC energies in the first row, the figure also shows results with the unscaled and scaled MP2 corrections in the second and third row respectively. The shaded region in each plot represents the error bar of the basis set extrapolation and contains contributions from both the current lattice constant and the subtracted reference value. Similar figures for LiH and Si can be found in the supplementary material\cite{supplement}. 

Several points are immediately clear from  \cref{fig:e_vs_a_mgo}. On the one hand, PBE orbitals not only have significant nonparallelity errors but they also decrease very slowly in magnitude with basis set size. Even though we decided not to pursue convergence of relative energy differences with PBE further in the case of MgO, it is apparent that over 1000 orbitals will be needed to reach convergence in the mHa range.
OPTM-GTO orbitals, on the other hand, show much more reasonable errors across the volume range and even small basis sets already produce acceptable results depending on the application.
In the case of MP2-NO orbitals we observe a somewhat erratic behavior in MgO, which is not observed in either LiH or Si. In the latter cases, both MP2-NO and OPTM-GTO orbitals produce comparable results and far superior to PBE in the case of LiH. At this point it is not clear what is the cause of the unusual behavior of MP2-NO orbitals in MgO as a function of volume, but it is clear that their performance is quite poor in this material. When MP2 corrections are applied to all three orbital sets, the magnitude of the errors are significantly reduced in all cases, particularly in the case of PBE orbitals. In this case, the scaled correction does not necessarily outperformed the unscaled version, even though it is usually superior on total energies. Notice that the scaling factor is calculated for every volume and system size independently. This implies that while the scaled correction leads to a more accurate estimate of the basis set incompleteness error at each volume, it does not mean that it's volume dependence is better. Nonetheless, both corrections seem to always provide a reduction in errors, of significant magnitude in some case. The picture in LiH and Si is consistent with the discussion above\citep{supplement} and with the results in \cref{fig:e_vs_N}.

\subsection{Performance and Cost Considerations}

Results presented in the previous sections indicate that the performance of each orbital set is heavily dependent on the system. PBE orbitals, for example, perform quite poorly on LiH and MgO, but are a good choice in Si. It is hard, in practice, to determine which orbital set is appropriate for a given system without more experience and testing. On the other hand, mean-field SCF orbitals like PBE have the smallest overhead associated with them and can be obtained from many widely available software packages. OPTM-GTO orbitals provide a good performance in all the systems we have considered to date, but have the additional burden of requiring an optimization framework and careful testing. Once the optimized orbital sets have been developed though, they represent a negligible overhead compared to the use of mean-field orbitals, since the evaluation of the periodic GTO orbitals on the FFT grid is a small overhead compared to the cost of evaluating the second-quantized Hamiltonian necessary for AFQMC and many other accurate orbital-based quantum many-body approaches. Out of the three orbital sets considered in this work, MP2-NO are the only ones that represent a significant overhead compared to the generation of the second-quantized Hamiltonian. Our current implementation of the MP2 energy scales roughly as $\mathcal{O}[n^2 m^2 N_k^3 N_g]$, where $\textit{n}$ is the number of electrons, $\textit{m}$ is the size of the basis set,  $N_k$ is the number of $\textit{k-points}$ and $N_g$ is the number of states in the auxiliary basis used to represent the electron repulsion integrals (typically given by the size of the FFT grid within the plane-wave cutoff for the density). The calculation of the MP2-NO is more expensive by roughly a factor of $\mathcal{O}{[m N_k]}$, which makes the evaluation of natural orbitals considerably more expensive than the evaluation of the energy. Even the evaluation of approximate natural orbitals for MP2, presented in Ref. \citep{GruneisNatOrb2011} as a reasonable alternative to full natural orbitals, represent an increased in cost of $\mathcal{O}{[m]}$ compared to energy evaluations. As the system size grows, this becomes significantly more expensive to the point that it becomes slower than the AFQMC calculation itself. This is mainly due to the fact that in MP2 calculations, we must keep all the states in the PW basis in order to produce high quality natural orbitals, so $\textit{m}$ is typically an order of magnitude larger in MP2-NO calculations compared to the number of states used in the subsequent AFQMC calculation. Taking this into consideration, we believe that the OPTM-GTO orbitals offer the best balance between performance and computational cost.

\subsection{Equation of State}

All results presented up to this point have been obtained from calculations on the primitive unit cell of the material. In this section we study the convergence of equation of state properties, including lattice constant, bulk modulus and atomization energies, on basis set and system size. We only present results for the OPTM-GTO basis on this section since it is the main focus on this work and since the calculation of MP2 natural orbitals represented a big computational cost in larger system sizes (more expensive than the AFQMC calculations themselves for systems with \textit{k-point} grids beyond $2\times2\times2$).

\begin{table}[htp]
\begin{center}
\begin{tabular}{*4c}
\hline
\hline
 {} & pVDZ & pVTZ & pVQZ \\
\hline
\hline
\multicolumn{4}{c}{Lattice Constant (Bohr)} \\
\hline
Uncorrected & 7.953(2)  &  7.921(2)  &  7.922(2) \\
MP2 corrected & 7.986(2)  &  7.964(2)  &  7.958(2) \\
Scaled MP2 corr. & 7.950(2)  &  7.947(2)  &  7.948(2) \\
\hline
\multicolumn{4}{c}{Bulk Modulus (GPa)} \\
\hline
Uncorrected & 195(2) & 178(2)  & 171(2) \\
MP2 corrected & 142(2)  & 165(2)  & 164(2) \\
Scaled MP2 corr. & 141(2)  & 163(2)  & 163(2) \\ 
\hline
\multicolumn{4}{c}{Atomization Energy (eV/atom)} \\
\hline
Uncorrected & & 4.932(2) & \\
MP2 corrected & 5.351(2) & 5.289(2) & 5.216(3) \\
Scaled MP2 corr. & 5.008(2) & 5.140(2) & 5.140(3) \\
\hline
\end{tabular}
\end{center}
\caption{Lattice constant, bulk modulus and atomization energy of MgO calculated with a $2\times2\times2$ \textit{k-point} grid (corresponding to a 16 atom simulation cell), for various OPTM-GTO basis sets. Results with and without basis set corrections are presented. }
\label{table:properties_kp2}
\end{table}%

\cref{table:properties_kp2} shows a summary of equation of state properties of MgO for various OPTM-GTO basis sets, for a Gamma-centered $2\times2\times2$ \textit{k-point} grid. We show results for \textit{uncorrected} AFQMC energies as well as for the 2 MP2-based corrections described above. Properties were obtained from a fit to the Vinet equation of state using results at seven volumes near equilibrium. Only the correlation energy was evaluated on the OPTM-GTO basis, the exchange energy was calculated separately with Quantum Espresso on a $10\times10\times10$ \textit{k-point} grid. Both basis set correction schemes give reasonable results at the pVTZ level already. We observe similar behavior in all three materials. See the supplementary material \cite{supplement} for a similar table with additional system sizes and for results on the other materials.

\begin{table}[htp]
\begin{center}
\begin{tabular}{*4c}
\hline
\hline
 {} & $E_{\mathrm{AFQMC}}$ & $E_{\mathrm{AFQMC}}^{\mathrm{Corrected}}$ & Exp \\
\hline
\hline
\multicolumn{4}{c}{Lattice Constant (Bohr)} \\
\hline
MgO & 7.914(3) & 7.900(6) & 7.916 \\
LiH & 7.556(3) & 7.558(3) & 7.519 \\
Si  & 10.261(2) & 10.234(6) & 10.244 \\
\hline
\multicolumn{4}{c}{Bulk Modulus (GPa)} \\
\hline
MgO & 178(2) & 170(4) & 169.8 \\
LiH & 39.3(7) & 38.4(7) & 40.1 \\
Si  & 101(1) & 106(3) & 100.8 \\
\hline
\multicolumn{4}{c}{Atomization Energy (eV/atom)} \\
\hline
MgO & 5.023(3) & 5.237(6) & 5.20 \\
LiH & 2.473(1) & 2.497(1) & 2.49 \\
Si  & 4.438(3) & 4.967(6) & 4.68 \\
\hline
\end{tabular}
\end{center}
\caption{Lattice constant, bulk modulus and atomization energy of MgO, LiH and Si calculated with both uncorrected and corrected AFQMC energies on a $3\times3\times3$ \textit{k-point} grid using a pVTZ OPTM-GTO basis set. Corrections include both finite-size effects as well as the scaled MP2 correction. Experimental results corrected for zero point effects (taken from Ref. \cite{schimka2011improved}) are also shown. }
\label{table:eos}
\end{table}%

We performed calculations on $2\times2\times2$, $3\times3\times3$ and $4\times4\times4$ \textit{k-point} grids and various OPTM-GTO basis sets. We used the results of the $3\times3\times3$ and $4\times4\times4$ \textit{k-point} grids to perform an extrapolation of the energies to the thermodynamic limit assuming a $1/N_k$ dependence, at each volume independently. We then combined these finite-size corrections with the scaled MP2-correction for basis set incompleteness to estimate the equation of state in the thermodynamic and complete basis set limits. \cref{table:eos} shows a summary of equation of state properties for all materials. We present results for \textit{uncorrected} AFQMC on a $3\times3\times3$ \textit{k-point} grid with a pVTZ basis, the same calculations with scaled MP2 basis set and finite-size corrections, as well as experimental results corrected for zero-point vibrational effects. Overall, we obtain excellent agreement with experimental results for almost all properties. Even \textit{uncorrected} AFQMC results are in reasonable agreement with experiment in most cases, which shows the excellent performance of the basis set generation scheme in these materials. We notice that the only significant discrepancy with experiment comes from the atomization energy of Si. While several effects can be contributing to this error, including: phase-less approximation in AFQMC, pesudo-potential approximations, and remaining finite-size error, a large fraction of the error is most likely coming from an overestimation of the basis set incompleteness error when using the scaled MP2 correction in Si. From \cref{fig:e_vs_N}, this overestimation is on the order of 0.14 eV in the primitive unit cell for the OPTM-GTO basis set. We are currently investigating alternative basis set correction schemes for systems where MP2 is not appropriate, including RPA-based methods. Results of this study will be presented in a future publication.

\section{Conclusions\label{sec:conclusions}}

We present an efficient approach for the generation of basis sets for correlated calculations in the solid state. By minimizing the MP2 energy of a small representation of the solid, we are able to generate a basis set that offers a convergence of the correlation energy as a function of system size that is comparable to more sophisticated approaches like the use of MP2 natural orbitals, but at a fraction of the cost. The resulting basis sets offer excellent transferability as a function of volume and lead to reasonably converged equation of state properties with modest sizes and much reduced computational costs. We also presented a simple but powerful basis set correction scheme based on MP2, inspired by recent work on coupled cluster theory \citep{IrmlerDuality2019}. The correction is able to provide a reasonable estimate of the basis set incompleteness error in AFQMC calculations and helps to extend the range of systems within range of the method.
We are currently exploring extensions of the work presented here, including the use of contracted basis sets, the use of RPA as the correlated method instead of MP2 and the transferability of the resulting basis sets with different environments and varying chemical composition. We believe that this work can pave the way for more robust and efficient basis set generation schemes in correlated solid state calculations, reducing one of the main hurdles in widespread application of quantum chemistry methods in solids.  

\begin{acknowledgments}
This work was supported by the U.S. Department of Energy (DOE), Office
of Science, Basic Energy Sciences, Materials Sciences and Engineering Division, as part of the Computational Materials Sciences
Program and Center for Predictive Simulation of Functional Materials (CPSFM). 
The work was performed under the auspices of the U.S. DOE by LLNL under Contract No. DE-AC52-07NA27344. An award of computer time was provided by the
Innovative and Novel Computational Impact on Theory and Experiment (INCITE) program. Some AFQMC calculations received computing support from the LLNL Institutional Computing Grand Challenge program.
\end{acknowledgments}

\bibliography{refs.bib}

\begin{thebibliography}{90}
\bibitem [{mat()}]{materials_data}%
  \BibitemOpen
  \href@noop {} {}\bibinfo {howpublished} {All simulation data is available at
  [url to be inserted].}\BibitemShut {Stop}%
\end{thebibliography}

\end{document}